\documentclass[lettersize,journal]{IEEEtran}
\usepackage{amsmath,amsfonts}
\usepackage{algorithmic}
\usepackage{array}
\usepackage[caption=false,font=normalsize,labelfont=sf,textfont=sf]{subfig}
\usepackage{textcomp}
\usepackage{stfloats}
\usepackage{url}
\usepackage{verbatim}
\usepackage{graphicx}
\def\BibTeX{{\rm B\kern-.05em{\sc i\kern-.025em b}\kern-.08em
		T\kern-.1667em\lower.7ex\hbox{E}\kern-.125emX}}
\usepackage{balance}
\usepackage{amsmath,amssymb,amsfonts}
\usepackage{graphicx}
\usepackage{algorithm}

\usepackage{tabularx} 
\usepackage{ragged2e} 

\usepackage{array}  
\usepackage{booktabs}   
\usepackage{rotating}
\usepackage{multirow}   
\usepackage{adjustbox}  

\usepackage{cite}  

\makeatletter
\newenvironment{breakablealgorithm}
{
		\begin{center} %
			\refstepcounter{algorithm}
			\hrule height.8pt depth0pt \kern2pt
			\renewcommand{\caption}[2][\relax]{
				{\raggedright\textbf{\ALG@name~\thealgorithm} ##2\par}
				
				\ifx\relax##1\relax
				
				\addcontentsline{loa}{algorithm}{\protect\numberline{\thealgorithm}##2}%
				
				\else
				
				\addcontentsline{loa}{algorithm}{\protect\numberline{\thealgorithm}##1}%
				
				\fi
				
				\kern2pt\hrule\kern2pt
				
			}
			
		}{
		
		\kern2pt\hrule\relax
		
	\end{center} %
	
}

\makeatother

\begin{document}

\title{LR-SQL: A Supervised Fine-Tuning Method for Text2SQL Tasks under Low-Resource Scenarios}
\author{ Wuzhenghong Wen , Yongpan Zhang , Su Pan , Yuwei Sun , Pengwei Lu and Cheng Ding
\thanks{Su Pan is with the College of Internet of Things, Nanjing University
	of Posts and Telecommunications, Nanjing 210003, China (e-mail: supan@
	njupt.edu.cn)}}

\markboth{}%
{}

\maketitle

\makeatletter
\renewcommand{\algorithmicrequire}{\textbf{Input:}}
\renewcommand{\algorithmicensure}{\textbf{Output:}}
\newcommand{\algorithmicbreak}{\textbf{break}}
\newcommand{\BREAK}{\STATE \algorithmicbreak}
\newcommand{\algorithmiccontinue}{\textbf{continue}}
\newcommand{\CONTINUE}{\STATE \algorithmiccontinue}

\makeatother

\begin{abstract}
Large language models revolutionize Text2SQL through supervised fine-tuning, yet a crucial limitation is overlooked: the complexity of databases leads to an increased context length, consequently resulting in higher GPU memory demands for model fine-tuning. To address this issue, we propose LR-SQL. LR-SQL comprises two supervised fine-tuning models: the schema\_link model and the SQL\_generation model, with the schema\_link model serving as the focal point for streamlining the overall process. During the fine-tuning of the schema\_link model, LR-SQL breaks down the complete database into flexible combinations of tables with adjustable quantities, enabling the model to learn the relationships within the entire database from these dispersed slices. Furthermore, to enhance the model's ability to perceive the relationships among various discrete slices during inference, LR-SQL trains the model's Chain-of-Thought capability for this task. Experimental results demonstrate that LR-SQL can reduce the total GPU memory usage by 40\% compared to existing fine-tuning methods, while only losing 2\% of table prediction accuracy in schema\_link task. For the overall Text2SQL task, the Execution Accuracy decrease by 0.6\%.Our project is now available on https://github.com/hongWin/LR-SQL
\end{abstract}


\section{Introduction}
\IEEEPARstart{T}{he} success of commercial large models in the market, notably GPT-4 \cite{achiam2023gpt} and PaLM2 \cite{anil2023palm}, has markedly accelerated the growth of the entire Large Language Model industry. This achievement has further catalyzed the advent of a multitude of open-source large models, including GLM-4 \cite{glm2024chatglm}, Qwen2\cite{yang2024qwen2}, and DeepSeek \cite{bi2024deepseek}. Due to industry requirements for privacy and low operational costs, utilizing instruction-tuned fine-tuning technology to transform open-source large language models into experts in their respective fields represents a new technological revolution in the field of AI applications. However, the issue of computational resources constitutes being the primary threshold for large language models to become domain-specific experts. Consequently, existing fine-tuning methods generally adopt Parameter-Efficient Fine-Tuning (PEFT) techniques. These techniques introduce partially trainable modules into each layer of the transformer structure of the large language model and only train these parameters during fine-tuning. This approach significantly reduces the number of parameters involved in training and decreases GPU memory usage. By utilizing PEFT technology, open-source language models can be trained under relatively modest resource constraints to serve as secure and efficient personal database assistants, with the core technology underpinning these assistants being Text2SQL technology. This technology is effective in addressing the demands for repetitive and relatively complex structured query language  (SQL) within practical production environments. \cite{liu2023comprehensive} indicates that medium-sized models, after undergoing supervised fine-tuning, exhibit superior performance compared to existing commercial large models in Text2SQL tasks , highlighting their strong application prospects. 

PEFT reduces the memory footprint for supervised fine-tuning of Text2SQL at the parameter level, but the length of the context input to the model is also one of the important factors affecting the memory usage during fine-tuning. it is important to note that GPU memory usage during fine-tuning is influenced by factors such as training batchsize which refers to the number of samples used in one training iteration of the model and the total number of tokens utilized for training. In the context of Text2SQL tasks, the structure and quantity of tables within the database play a pivotal role in determining the overall number of tokens required for the task. As the number of tables and columns increases, the model needs higher number of tokens to encode these features, ultimately resulting in escalated GPU memory requirements during the supervised fine-tuning process.In practical application scenarios, large databases frequently encompass hundreds of tables. When employing the methods outlined in \cite{pourreza2024dts}, \cite{xue2023db}, \cite{shen2023spsql} to directly establish relationships among query requirements, query predictions, and the comprehensive database information, a substantial number of tokens are necessitated for encoding per instance during the supervised fine-tuning phase. Consequently, given the constraints of limited GPU memory resources, it becomes infeasible to carry out supervised fine-tuning on the mode.

The currently supervised fine-tuning method improved based on PEFT and designed to further reduce memory usage during fine-tuning, cannot be adapted and applied to large language models of all architectures.Current advancements in PEFT, such as QLORA \cite{dettmers2024qlora}, are capable of reducing the memory requirements for fine-tuning by decreasing the precision of model parameters. Long LoRA \cite{chen2023longlora}, on the other hand, can both extend the context window of LLAMA2 \cite{touvron2023llama} and further decrease the memory footprint of supervised fine-tuning at the token level. However, both of these methods require the design of specific implementation codes tailored to large language models with different architectures, necessitating a deep understanding of the model's underlying mechanics. Therefore, in practical applications, these techniques cannot be rapidly generalized to large models with diverse architectures.To our best knowledge , we find that no prior work has focused on employing supervised fine-tuning methods to reduce memory consumption in Text2SQL tasks.

Hence, we make improvements to the supervised fine-tuning objectives for the Text2SQL task, thereby reducing the memory consumption during supervised fine-tuning at token level. Specifically, we group all tables in the database into several slices based on specific relationships, enabling these slices to flexibly cover varying numbers of tables according to GPU memory constraints, which controls the number of tokens input to the model. However, due to the disruption of database integrity, directly constructing the relationship from the problem to the final SQL using the method from [16] is infeasible. To enable the model to capture the relationship between query requirements, the target SQL, and the entire database from dispersed database table information (slices), We decompose the entire Text2SQL task into two distinct sub-tasks: one is defined as the schema linking task, which aims to predict the tables relevant to the problem, and the other is the SQL generation task, responsible for generating the query statements relevant to the problem. During the fine-tuning phase of the schema linking task, we train the model to establish relationships between the problem, the entire database, and the target tables from dispersed slices. In the inference phase of the schema linking task, we inquire the model about the relationship between the problem and each slice successively and integrate the results of multiple inquiries to derive all table names relevant to the problem. This approach allows the model to recover its understanding of the entire database from the problem and dispersed slices, and to predict the table names in the output SQL queries as accurately as possible. Although the aforementioned schema linking task effectively mitigates the memory issue during training, it severs the connections between slices during inference, preventing the model from effectively utilizing previous inference information to aid in subsequent decisions. Since our designed schema linking task produces results through multiple inquiries during inference, the Chain-of-Thought (CoT) approach \cite{wei2022chain} can endow the model with human-like thinking to guide the model step-by-step towards generating the final result. Therefore, we integrate our inference method with the  approach, incorporating the table information inferred by the model in previous steps into the known information for current inference. However, the CoT approach is a form of prompt engineering that stimulates the model's knowledge potential through specifically designed thinking templates to solve the final answer through multi-step inference. As our schema linking task involves the injection of new domain knowledge by splitting the database into multiple slices for multi-step inference, merely integrating the CoT method during inference is insufficient for the model to perceive the gap from previous inference information to the optimal solution. We need to inject new database knowledge into this logical chain of CoT, combining the model's logical thinking with the new database knowledge. Therefore, during inference, we sequentially inquire the model about the connections between slices and the related problem, and integrate previously inferred information into the current inference. During the supervised fine-tuning phase, we incorporate the optimal target tables generated from previous slices into the training information for the current step, following the order defined during inference. Based on this, while reducing memory usage during supervised fine-tuning in the schema linking phase, we enable the model to have a more macroscopic and global perspective among dispersed slices.After filtering out a large number of irrelevant tables using the schema linking model, the predicted target tables and problems are sent to the SQL generation model to generate the final SQL. This model is fine-tuned using only a few tables that contain the target tables, significantly reducing the required memory. Our method, LR-SQL(A supervised fine-tuning method for Text2SQL tasks under low-Resource Scenarios), has been extensively tested on two new datasets constructed from the Spider \cite{yu2018spider} dataset using different open-source models. The experimental results demonstrate that LR-SQL can effectively reduce the memory required for fine-tuning compared to the baseline, while maintaining similar performance across all evaluation metrics. Our contributions are summarized as follows:

(1)We develop a complete supervised fine-tuning method for Text2SQL under low GPU memory Scenarios, which include schema linking model and  SQL generation model. Meanwhile, we constructed two databases based on the Spider dataset, comprising 254 and 172 tables respectively, to simulate Text2SQL tasks in large-scale and medium-scale database scenarios, and subsequently developed two corresponding sets of datasets to simulate these scenarios. We conducted extensive experiments on the constructed datasets using different models, and demonstrated that LR-SQL can effectively reduce memory usage during supervised fine-tuning while maintaining performance similar to the baseline across different metrics. 

(2) Under resource-constrained scenarios, we propose an innovative schema linking task that decomposes the database into multiple slices with adjustable token capacities, allowing the model to learn the relationship between questions and the database in segments. To enable the model to have a more macroscopic and global perspective among dispersed slices during inference, we incorporate the capabilities of the Chain-of-Thought through supervised fine-tuning.

\section{Reference Work}
Text2SQL can be classified as a pipeline method that consists of two primary components: schema linking and SQL generation. The schema linking component is dedicated to establishing relationships between questions, databases, and target tables or columns, whereas the SQL generation component is responsible for generating SQL queries that are concerning to the given question, based on the provided question and relevant tables. Existing research on Text2SQL can be classified into three main categories: 

\subsection{Encoder-Decoder Model Base}
This method designs an appropriate encoder to capture the relationship between questions and SQL syntactic structures, tailored to the grammatical characteristics of SQL statements, and devises a set of decoder paradigms for effective decoding. Specifically, \cite{xiao2022cqr} employs auxiliary conversational question reconstruction (CQR) learning to capture context dependencies and designs an encoder model aimed at enhancing the SQL generation capability of large language models. This approach efficiently extracts valuable information from historical context, guiding the model to generate more standardized answers. Furthermore, \cite{zheng2022hie} addresses the discrepancy between natural language and the logical form of SQL language by encoding SQL queries and natural language contexts in a multimodal manner during the pre-training phase, effectively improving the SQL hit rate of pre-trained language models. Additionally, \cite{hui2022s} proposes a novel encoder that encodes grammatical information, linking structures, and schema information, while introducing decoupling constraints to enhance model performance. However, it is worth noting that models designed with specific task-oriented encoder-decoder architectures may exhibit reduced generalization capabilities when applied to other tasks.	

\subsection{Zero-shot or Few-shot Learning Based on Large Language Models}
This method employs a zero-shot or few-shot approach, carefully designing appropriate prompt templates and inference processes to extract the inherent understanding and judgment capabilities of language models. Specifically, \cite{lee2024mcs} utilizes multiple sets of prompt templates to expand the search space of language models and designs a selector to choose the optimal SQL. This approach can effectively enhance the likelihood of language models deriving the optimal answer. Furthermore, \cite{pourreza2024din} designs an external plugin based on the method of Chain-of-Thought to enhance the ability of large language models to generate target SQL. Experimental results demonstrate that this framework can effectively improve the accuracy of SQL generation and possesses error-correction functionality for the generated SQL. Additionally, MAC-SQL\cite{wang2023mac} utilizes GPT4 to design a complete collaborative framework for generating SQL statements, which includes: 1) table and column selectors, 2) question decomposers, and 3) SQL rewriters. Lastly,  C3\cite{dong2023c3} designs a comprehensive Text2SQL inference framework for GPT4 in zero-shot scenarios, which include prompt template design, prompt calibration, and inference result selection. These methods stimulate the domain knowledge of language models through specific approaches. However, if the model lacks knowledge in a particular domain, fine-tuning may be necessary to supplement the model with vertical domain knowledge and align it with human preferences.

\subsection{Supervised Fine-tuning Based on Large Language Models}
Supervised fine-tuning represents a pivotal approach aimed at addressing the specific requirements of vertical domains by compensating for the specialized knowledge deficits in pre-trained models, thereby demonstrating robust adaptability to novel data. \cite{pourreza2024dts} The Text2SQL task is segmented into two distinct sub-tasks: schema linking and SQL generation. Subsequently, the pre-trained language model undergoes fine-tuning with supervision, grounded in these two sub-tasks. This methodology refines the prompt fine-tuning template, eliminates redundant columns, and ultimately enhances the accuracy of SQL execution. \cite{sun2023instruction} A specialized SQL question-answering model is trained using the China State Grid SQL question-answering dataset. This method leverages a large model fine-tuned with professional knowledge in the power grid domain as its foundation and incorporates Text2SQL question-answering data for instruction-based supervised fine-tuning. Experimental results underscore that fine-tuning with specialized knowledge and SQL question-answering data in professional domains can effectively augment the model's SQL generation performance within those domains. \cite{xue2023db} Through the integration of RAG technology, the performance of Text2SQL is bolstered in both professional and private domains, providing a comprehensive baseline for model supervised fine-tuning. This method amalgamates multiple pipeline technologies to develop a holistic Text2SQL system that can be deployed offline locally, effectively safeguarding local privacy.  SC-Prompt\cite{gu2023few} outlines two sets of downstream tasks for supervised fine-tuning: SQL structure generation and content filling. The SQL structure generation task designed within this method exhibits greater effectiveness in learning from expert experience and knowledge. The task of SPS-SQL\cite{shen2023spsql} is segmented into table selection, column selection, SQL generation, and value filling, utilizing four models to sequentially generate the final SQL. \cite{li2024table} A novel paradigm of "table tuning" is introduced, designing diversified table prediction tasks to enhance the language model's table and column prediction capabilities. DELLM\cite{hong2024knowledge} utilizes reinforcement learning to refine the language model after supervised fine-tuning. This method facilitates continuous improvement in model performance through the interaction between the system and the model.

The aforementioned supervised fine-tuning methodologies have been successfully employed for schema linking and SQL generation tasks within the confines of small databases, adeptly executing instruction-supervised fine-tuning on a solitary GPU. Nonetheless, when confronted with medium-to large-scale databases including numerous tables, the enormous context included by each instance  demand considerable GPU memory allocation for training purposes. Consequently, the primary focus of our paper is to devise a solution for executing the schema\_link task within large databases while operating under constrained computational resources.

\section{Method overview}
\subsection{Task description of Text2SQL}
The goal of Text2SQL is to establish a critical path from query questions to structured query language statements, and can be defined as:

\begin{equation}  
F\left( {Q,S} \right) \to A \label{eq:3.1}  
\end{equation}

where $Q$ is the set of query questions, $S$ is the description of the database, and $A$ is the SQL query statement predicted by the model. Furthermore $S \in \{ T_1, T_2, \ldots, T_n, \ldots, T_N \}$ ,  where $T_n$ represents the information of the $n$-th table within the database, characterized by the relationship$\{{c_{n1}},{c_{n2}},\ldots,{c_{ni}},\ldots,{c_{nI}};c_{n1}',\ldots,c_{nj}',\ldots,c_{nJ}'\}$, where $c_{ni}$ denotes column $i$ in $i$-th table, while $c_{nj}'$represents the $j$-th foreign key which refers to one or more fields in a table, where the values of these fields must correspond to the values of the primary key or unique key in another table, hence the $I$ and $J$ have the following constraints with $I>0,J\geq0$.
\subsection{The Application of Supervised Fine-Tuning Methods in Text2SQL}
Supervised fine-tuning technique is a model training method that enables the model to follow input instructions and think and respond according to a predetermined template. Taking GLM4\cite{glm2024chatglm} as example, the construction of a single training data in the Text2SQL task is illustrated in Fig. 1:
\begin{figure}[htbp]
	\centering
	\includegraphics[scale=0.5]{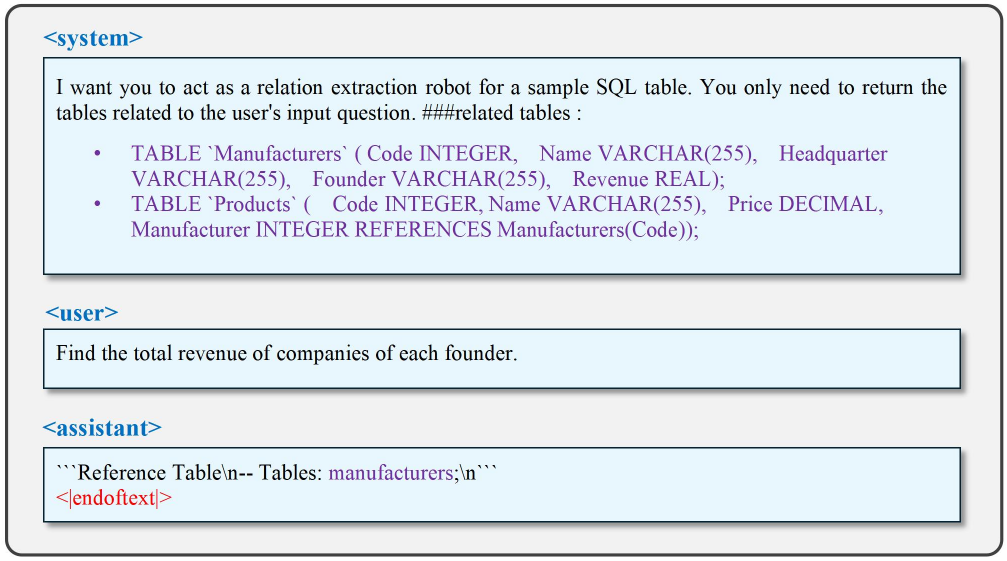}
	\caption{Supervised Fine-Tuning of Templates in Text2SQL Tasks}
\end{figure} 

In Fig. 1, $<\text{system}>$, $<\text{user}>$, and$<\text{assistance}>$ are role labels used to inform the model of the responding roles during inference. Specifically, the content following $<\text{system}>$ indicates the instructions that the model needs to follow, while the content after $<\text{user}>$ conveys the current user's demands to the model. The content following $<\text{assistance}>$ represents the model's expected output, which can also be defined as the label for the model's prediction. During supervised fine-tuning, the model predicts the content after $<\text{assistance}>$ based on the content in both $<\text{assistance}>$ and $<\text{user}>$, and then compares it with the label to compute the loss function. The $<\text{endoftext}>$ token serves as an end marker to prevent the model from diverging in thought and thereby reducing inference time in subsequent reasoning stages. By defining a supervised fine-tuning template, the model can evoke fine-tuned knowledge according to the template during inference and the users can easily extract answers from the pre-established response template. 

Recent research focused on supervised fine-tuning techniques for Text2SQL, two fundamental tasks are encompassed: schema\_linking and SQL generation. The objective of the schema\_link task can be explained by (2) :
\begin{equation}  
	\begin{aligned}  
		& \underset{\theta}{\text{minimize}}  
		& & \frac{1}{T} \sum_{i=1}^{T} \text{Loss}(s^{*}_i, \text{LLM}_S(q_i, S, D; \theta))  
	\end{aligned}  
\end{equation}

(2) means to minimize the empirical loss, enabling the schema\_link large language model $LLM_S$  to predict tables and table contents ${s^{*}}_{i}$ that match the SQL query requirements $q_{i}$ , given the database $D$ and all table descriptions $S$ in the database. $\theta$ represents all parameters involved in the training, and $Loss$ is the loss function used in the supervised fine-tuning process.

The objective of the SQL generation task can be illustrated by (3):
\begin{equation}
	\begin{aligned}
		& \underset{\theta}{\text{minimize}}
		& & \frac{1}{T} \sum_{i=1}^{T} \text{Loss}(\text{GOLD}_i, \text{LLM}_G(q_i, S, D; \theta)) 
	\end{aligned}
\end{equation}

(3) enables the large language model for SQL generation to accurately predict SQL queries that are as close as possible to the "Expected SQL Query" which we defined as $\text{GOLD}_{i}$, given a database $D$, descriptions of all tables $S$ in the database, and the specific SQL query demand $q_{i}$. $\text{LLM}_{G}$ denotes the large language model specifically designed for SQL generation, while $\theta$ represents the parameters involved in the training process. Furthermore, if the table names and its corresponding schema which represents the complete table structure information are predicted during the schema\_link stage, (3) can be appropriately rewritten as follows:
\begin{equation}  
	\begin{aligned}  
		& \underset{\theta}{\text{minimize}}  
		& & \frac{1}{T} \sum_{i=1}^{T} \text{Loss}(\text{GOLD}_i, \text{LLM}_G(q_i, (s'_i \cup s^*_i), D; \theta))  
	\end{aligned}  
\end{equation} 

Where $({s'}_{i} \cup {s^{*}}_{i})$ denotes integrating the table prediction results derived from the schema\_link model with the tables concerning to the target SQL during the supervised fine-tuning phase. The primary objective of this integrating is to augment the generalization and error correction capabilities of the SQL generation model compared to the method take ${s^{*}}_{i}$ only.

The SQL Generation task outlined in (3), identified as \cite{xue2023db}, establishes a direct relationship between the input question and the database, ultimately generating the corresponding SQL query. Typically, (2) and (4) are arranged in a pipeline format, designated as \cite{pourreza2024dts}\cite{shen2023spsql}. In this pipeline, the schema\_link model is initially employed to filter out potential tables and columns. Subsequently, the filtered content, denoted as ${s'}_{i}$, is relayed to the SQL generation model for further processing.

\section{Supervised fine-tuning methods for schema\_link task under low-resource conditions}
Both the schema\_link task and the SQL Generation task incorporate all table descriptions $S$. In existing supervised fine-tuning methodologies, several factors influence GPU memory usage during the fine-tuning process, including the model's parameter count, the batchsize utilized for training, the extent of model quantization, and the total number of tokens in each training data instance. Notably, in Text2SQL tasks, the comprehensive table description $S$ of the database serves as the primary determinant of the total number of tokens in each training data instance.

In current supervised fine-tuning practices, $S$ is positioned after  \textquotedblleft\#instruction\#\textquotedblright in Figure 1, following which the entire supervised fine-tuning template is encoded to construct the training data, which is dependent on the size of $S$. In approaches such as \cite{pourreza2024dts}, \cite{xue2023db} and \cite{shen2023spsql}, where the size of $S$ is comparatively small, supervised fine-tuning can be accomplished on a professional GPU by adjusting the batchsize for training after encoding the training data. However, in scenarios involving hundreds of tables, encoding single training instance can yield thousands or even tens of thousands of tokens, making it challenging for the condition under low GPU memory. Consequently, summarizing the relationship between questions, databases, and SQL queries in a single step through supervised fine-tuning is not practical (3).

One of the primary tasks of schema\_link is to extract tables ${s^{*}}_{i}$ from $S$ that are highly concerning to the question. Despite the large size of $S$, there exists significant discreteness between tables and weak contextual relationships. Therefore, we make improvements to (2) by decomposing  into slices. These slices can dynamically control the number of tables, which in turn determines the number of tokens input into the model. Subsequently, we devise a novel approach to solve the table prediction problem, enabling the model to identify whether the tables relevant to the query exist within these slices. During inference, the query interacts with these slices multiple times, ultimately retrieving the table names relevant to the query from a large number of tables.The entire methodology is depicted in Fig. 2.


\begin{figure}[htbp] 
	\centering 
	\includegraphics [scale=0.55]{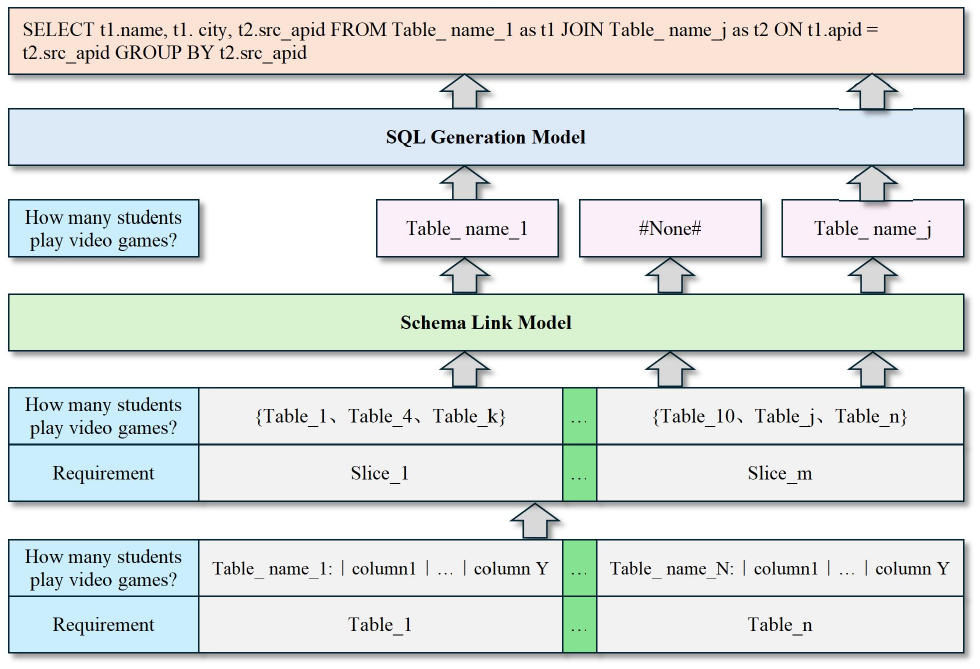}
	\caption{The overall framework of LR-SQL}
\end{figure}

\subsection{Disassembling Database Data}
the rigorous associations between tables within a database are typically governed by foreign key constraints. Often, multi-table query requirements need aggregating data from several tables that are interrelated through these foreign keys to derive answers. By identifying and using the foreign key relationships among tables, can finally enhance the performance of multi-table prediction tasks in Text2SQL. Therefore, the primary step of this work involves establishing initial relationships between tables based on foreign key associations. We categorize the table compositions within large databases according to the partition defined in (5).
\begin{equation}  
S = S_{\text{Reference}} \cup S_{\text{no\_Reference}}  
\end{equation}

As depicted in Equation (5), we categorize all tables in the database into two distinct groups. The group $S_{\text{Reference}}$ denotes the collection of tables that are interconnected through foreign keys, with the relationship expressed as $S_{\text{Reference}} = \{ G_1, G_2, \ldots, G_i, \ldots, G_I \}$. $G_i$ signifies a correlation group, implying that every table within this group is subject to constraints imposed by the $i$-th foreign key or its related foreign keys. We further define $G_i = \{ t_{i1}, t_{i2}, \ldots, t_{ij}, \ldots, t_{iJ} \} $, where $t_{ij}$ encapsulates the description of the table name and all its columns within the $i$-th correlation group. This is elaborated by the reference $t_{ij} \in \{ c_{ij1}, c_{ij2}, \ldots, c_{ijk}, \ldots, c_{ijK}; c_{ij1}', \ldots, c_{ijv}', \ldots, c_{ijV}' \}$, where $c_{ijk}$ represents the $k$-th column in the $j$-th table of the $i$-th correlation group, and $c_{ijv}'$ denotes the $v$-th foreign key correlation with $v > 0$ . The second group is $S_{\text{no\_Reference}}$, which represents the set of tables that without any foreign key correlations, with the relationship ${S_{no\_{\mathop{\rm Re}\nolimits} ferce}} = \{ {t_1},{t_2},..,{t_o},..,.{t_O}\}$ . Here ${t_o}$, to denotes tables without foreign key correlations and all their respective columns, which is further defined as ${t_o} \in \{ {c_{o1}},{c_{o2}},..,{c_{op}},..,{c_{oP}}\}$ , where $c_{op}$ represents the p-th column in the $o$-th table.

\subsection{The method for low-resource training in schema\_link task }

Although  we  decompose $S$ into its constituents $S_{\text{Reference}}$ and $S_{\text{no\_Reference}}$ , the token generated by encoding these components using the $Tokenizer()$ function during the training phase may still massive. As a result, a more detailed disassembly of $S_{\text{Reference}}$ and $S_{\text{no\_Reference}}$ becomes necessary. We now define the minimal unit slice $w$ of table information in supervised fine-tuning, as illustrated in Fig. 3 To clarify, $max\_token$ denotes the maximum token capacity of a single data instance when the model is fine-tuned with one batchsize under the memory capacity of $\chi$ . Meanwhile, $model\_token$ represents the token count for encoding the model's special characters and the supervised fine-tuning template, as highlighted in black font in Fig. 1 Consequently, the total token count $slice\_token$, following the encoding of $w$, must meet the condition $slice\_token < \max \_token{\rm{ }} - {\rm{ }}model\_token$. The procedure for transforming $S$ into the set of total slices which define as $W$ is described in detail in Algorithm 1.

\begin{breakablealgorithm}
	\caption{Constructing The Slice Set $W$}
	\begin{algorithmic}[1] 
		\REQUIRE $slice\_token$, $S_{\text{Reference}}$, $S_{\text{no\_Reference}}$.
		
		\ENSURE Slice Set $W$.
		
		\STATE Initialize $w \leftarrow \{ \}$, $W \leftarrow \{ \} $
		
		\WHILE{$G \in {D_{{\mathop{\rm Re}\nolimits} ferce}}$}
		
		\WHILE{$g \in G$}
		
		\IF{$len(Tokenizer(g) + Tokenizer(w)) < slice\_token$}
		
		\STATE $w.add(g)$;
		
		\ELSE 
		
		\STATE $W.add(w)$, $w \leftarrow \{ \}$;
		
		\STATE $w.add(g)$;
		
		\ENDIF
		
		\ENDWHILE
		
		\ENDWHILE
		
		\WHILE{$T \in {D_{{\mathop{\rm Re}\nolimits} no\_ferce}}$}
		
		\IF{$len(Tokenizer(T) + Tokenizer(w)) < slice\_token$}
		
		\STATE $w.add(T)$;
		
		\ELSE 
		
		\STATE $W.add(w)$, $w \leftarrow \{ \}$;
		
		\STATE $w.add(T)$;
		
		\ENDIF
		
		\ENDWHILE
		
		\STATE Return $W$;
		
	\end{algorithmic}
\end{breakablealgorithm}

Through Algorithm 1, we decompose $S$ into $S_{\text{Reference}}$ and $S_{\text{no\_Reference}}$, and subsequently populate slices with the table contents from these components, adhering to a predefined maximum slice capacity. This transformation results in $S$ being converted into a set $W$, which comprises multiple slices denoted as $w$. Each slice $w$ acts as the smallest, distinct unit of table information during the supervised fine-tuning process. In contrast to conventional schema\_link task approaches, once $S$ is decomposed into $W$, we must progressively reassemble and recover the entire database's table information from these slices. As such, our schema\_link task need multiple iterations of target extraction from the various slices $w$. The detailed procedure of our schema\_link task is outlined as follows:
	\begin{equation}  
	\begin{aligned}  
		& \underset{\theta}{\text{minimize}}  
		& & \frac{1}{NM} \sum_{i=1}^{N} \sum_{j=1}^{M} \text{Loss}\left(s_j^*, \text{LLM}_S(q_i, W_j; \theta)\right) \\  
		& \text{subject to}  
		& & \text{len}(Tokenizer(w_j)) < slice\_token ,\\  
		& & & S_i^* = \sum_{j=1}^{M} s_j^*  
	\end{aligned}  
\end{equation}

Where $N$ represents the overall count of questions contained within the dataset. $w_j$ stands for the $j$-th slice in total slice set $W$, where $S = W = \sum\nolimits_{j = 1}^M {{w_j}}$, and $M$ indicates the total number of slices. ${s^*}_j$ denotes the target table names for slice $w_j$, if no table concerning to the question are found within that slice, it is assigned the special token "\#None\#". Once predictions for all slices related to question $q_i$ are complete, a definitive table prediction result $S'$ is derived.

\begin{figure}[h]   
	\begin{minipage}{\textwidth} 
		\includegraphics [scale=0.65]{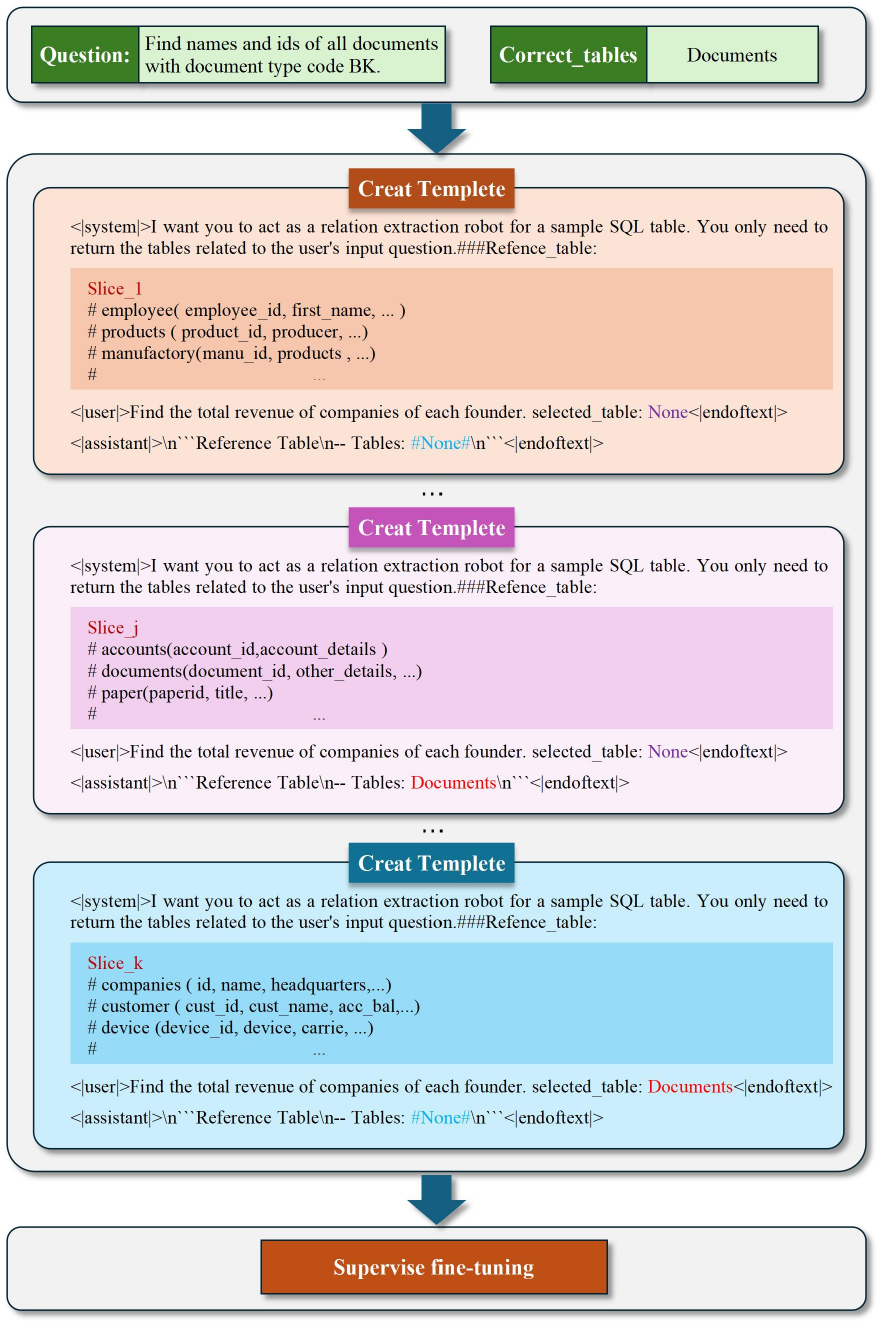}  
		\caption{Preparing training data for LR-SQL}   
	\end{minipage}  
\end{figure}

However, when using (6) to achieve the task objectives, the model faces a challenge: it can only identify tables potentially relevant to the query within each slice based on current information during the reasoning process. Nevertheless, the table information inferred from previous slices may contribute to partial solutions for the overall problem, indicating that earlier conclusions could positively impact future decisions. The reasoning method that incrementally breaks down the problem and combines the reasoning results at each step is known as Chain-of-Thought \cite{wei2022chain}. While Chain-of-Thought stimulates the model's inherent knowledge, the model lacks relevant knowledge about the private database before fine-tuning. Therefore, during supervised fine-tuning, we instill database knowledge into the model along with Chain-of-Thought's reasoning capabilities, allowing the model to recognize short-term connections between the current slice and previous slices, as well as long-term relationships between each discrete slice and the original intact database. Fig .3 illustrates the supervised template construction for LR-SQL, where we first define the reasoning sequence of slices and then integrate the table information inferred from previous slices into the $<\text{user}>$ context to aid decision-making at each step. As the sample size increases during fine-tuning, the model will memorize this reasoning sequence and incorporate prior reasoning information during inference. Through Experiment 5-D-4), we show that incorporating the Chain-of-Thought mindset into the model ultimately improves the final table prediction performance. Consequently, our task objective is transformed into (7).

\begin{equation}  
	\begin{aligned}  
		& \underset{\theta}{\text{minimize}}  
		& & \frac{1}{NM} \sum_{i=1}^{N} \sum_{j=1}^{M} \text{Loss}\left(s_j^*, \text{LLM}_S(q_i, W_j, s_{0,\ldots,j-1}^*; \theta)\right) \\  
		& \text{subject to}  
		& & \text{len}(Tokenizer(w_j)) < slice\_token ,\\  
		& & & S_i^* = \sum_{j=1}^{M} s_j^*  
	\end{aligned}  
\end{equation}  

Based on our final equation (7), for the same problem $i$, we integrate the target table information $s_{0,..,j - 1}^*$ covered by the first $j$-1 slices into the training objective of the $j$-th slice. By constructing the optimal objectives for different problems across different slices, we enable the model to learn the relationship between slices and problems while injecting the Chain-of-Thought reasoning approach.The training process of LR-SQL is presented in Algorithm 2, where $F_{template}$ denotes the template construction function. This function is utilized to insert information such as the problem, slices, and prediction targets into the supervised fine-tuning template illustrated in Fig. 3.

\begin{breakablealgorithm}
	\caption{LR-SQL Training}
	\begin{algorithmic}[1] 
		\REQUIRE $\text{Schema\_link Task Dataset}\hspace{5pt} Dataset\_QA$, $\text{Slicing Set}\ W$, $\text{training step}\ T$.

		\STATE $Model$ = Initialize\_model()
		
		\STATE $Optimizer$ = AdamW(model.parameters())
		
		\WHILE {$\text{step} < T$}
		
		\STATE $loss.Initialize()$
		
		\WHILE{$Dataset\_QA(q,{{\rm{S}}^*})$}
		
		\STATE $Selected \leftarrow \{ \}$
		
		\WHILE{$({w_0},...,{w_i},...{w_N}) \in W$}
		
		\WHILE{${s^*} \in {S^*}$}
		
		\IF{${s^*}$ in $w_i$}
		
		\STATE ${y_{pre}} = Model({F_{template}}(q,{w_i},Selected);\theta )$;
		
		\STATE ${y}_{\_pre} = Model({F_{template}}(q,{w_i});\theta )$; //Balancing the number of positive samples 
		
		\STATE $loss{\rm{ }} + = {\rm{ }}Loss\_Func({y_{pre}},{s^*})$;
		
		\STATE $loss{\rm{ }} + = {\rm{ }}Loss\_Func({y}_{\_pre},{s^*})$;
		
		\STATE $Selected.add({s^*})$;
		
		\ELSE 
		
		\STATE ${y_{pre}} = Model({F_{template}}(q,{w_i},Selected);\theta )$;
		
		\STATE $loss{\rm{ }} + = {\rm{ }}Loss\_Func({y_{pre}},\text{"None"})$;
		
		\ENDIF
		
		\ENDWHILE
		
		\ENDWHILE
		
		\ENDWHILE
		
		\STATE $loss.backward()$;
		
		\STATE $Optimizer.step()$;
		
		\ENDWHILE
			
	\end{algorithmic}
\end{breakablealgorithm}

\subsection{Model inference}
\begin{figure}[h]   
	\begin{minipage}{\textwidth} 
		\includegraphics [scale=0.5]{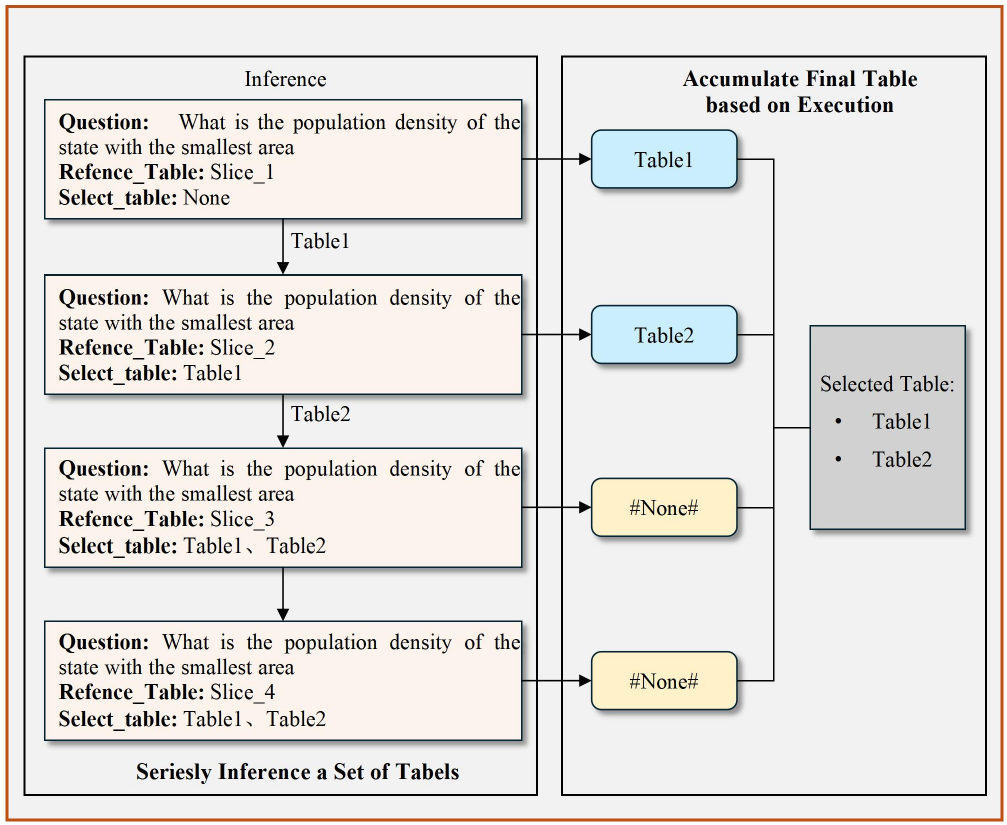}  
		\caption{Model inference process}   
	\end{minipage}  
\end{figure}

After fine-tuning the model for the task designed in accordance with (7), the table name prediction task is transformed into a multi-step prediction process. Prior to conducting inference, it is necessary to construct a set of segments. The flowchart illustrating the inference process is depicted in Fig. 4 Notably, during the inference phase, the memory requirements for long  texts are substantially reduced compared to those during training. Consequently, when performing model inference, we can rigorously adhere to the settings outlined in Algorithm 2, where the size of each segment $\_w$ equal to The detailed inference procedure is presented in Algorithm 3.

After fine-tuning the model for the task designed in accordance with (7), the process of solving the schema\_link task is transformed into a sequence of reasoning steps using the Chain-of-Thought (CoT) approach to infer potential table names from multiple slices, as illustrated in Figure 4. Notably, during the inference stage, the memory required for the model to process long sequence texts is significantly reduced compared to the training stage. Therefore, when performing model inference, we can strictly adhere to the settings outlined in Algorithm 2, setting the size of each slice $w$ equal to $slice\_token$. The inference process of LR-SQL is presented in Algorithm 3.
\begin{breakablealgorithm}
	\caption{Model Inference Process}
	\begin{algorithmic}[1] 
		\REQUIRE $\text{Query question}\ Q$, total tables $S$ under target database $D$.
		
		\ENSURE $\text{Set of relevant tables}\ {S'}$. 
		
		\STATE $\_W \leftarrow {F_{slice}}(S);\text{//algorithm 1}$
		
		\WHILE{$\_w \in \_W$}
		
		\STATE ${s'} \leftarrow LLM({F_{Model}}(Q,\_w,{S'}))$;
		
		\STATE ${S'}.add({s'})$;
		
		\ENDWHILE
		
		\STATE Return ${S'}$;
	\end{algorithmic}
\end{breakablealgorithm}

As can be seen from the inference algorithm, our method achieves the goal of reducing the memory required for training by configuring the size of the slice $w$. However, this increase the number of inference iterations necessary to complete the entire schema\_link task. During inference, varying the size of $\_w$ impacts not only the number of iterations needed to arrive at the final result but also the precision of that result. The experiments examining the relationship between different slice sizes $\_w$ during the inference process and the ultimate inference accuracy will be detailed in Chapter 5.

\section{EXPERIMENTS}
\subsection{Dataset}
Introduced in 2018, Spider\cite{yu2018spider} is a premier Text2SQL dataset, featuring 164 databases and 9693 natural language . It classifies SQL queries into four difficulty levels (easy, medium, hard, extra hard) based on components and conditions, and innovatively introduces Exact Matching and Execution Accuracy as evaluation metrics for Text2SQL tasks.
\subsection{Experimental model and hyperparameter settings}

\begin{table*}[htbp]  
	\centering  
	\caption{Performance Evaluation of Different Models for the Table Prediction Task in the Schema\_link }  
	\begin{tabular}{@{}c@{\hspace{4pt}}c@{\hspace{4pt}}c@{\hspace{4pt}}c@{\hspace{4pt}}c@{\hspace{4pt}}c@{\hspace{4pt}}c@{}}  
		\toprule  
		\multicolumn{7}{c}{\textbf{Dataset1}} \\  
		\midrule  
		Method & \centering Total \newline Accuracy & Filtered \newline Accuracy & Average \newline Precision & Average \newline Recall & GPU usage (MB) & Inference time(s) \\  
		\midrule  
		GLM4 (zero-shot) & 20.62 & 66.58 & 54.0 & 76.97 & 21333 & 1.12 \\  
		Qwen2 (zero-shot) & 9.30 & 13.84 & 27.62 & 21.08 & 32884 & 1.39 \\  
		DeepSeek (zero-shot) & -- & -- & -- & -- & Out of Context Window & -- \\  
		GLM4 (compromise) & 78.59 & 83.81 & 87.89 & 88.00 & 59472 & 1.02 \\  
		Qwen2 (compromise) & 72.84 & 78.85 & 88.60 & 87.02 & 33736 & 1.23 \\  
		DeepSeek (compromise) & 60.57 & 70.50 & 83.14 & 81.59 & 67770 & 0.62 \\  
		GLM4 (LR-SQL) & 91.38 & 94.26 & 95.50 & 95.76 & 57152 & 6.49 \\  
		Qwen2 (LR-SQL) & 84.07 & 87.21 & 92.78 & 91.92 & 48310 & 5.49 \\  
		DeepSeek (LR-SQL) & 79.37 & 86.16 & 91.31 & 91.23 & 77650 & 8.82 \\  
		GLM4 (DTS-SQL) & -- & -- & -- & -- & OOM & -- \\  
		Qwen2 (DTS-SQL) & -- & -- & -- & -- & OOM & -- \\  
		DeepSeek (DTS-SQL) & -- & -- & -- & -- & Out of Context Window & -- \\
		\midrule  
		\multicolumn{7}{c}{\textbf{Dataset2}} \\  
		\midrule  
		& \centering Total \newline Accuracy & Filtered \newline Accuracy & Average \newline Precision & Average \newline Recall & GPU usage (MB) & Inference time(s) \\  
		\midrule  
		GLM4(zero-shot) & 25.84 & 74.72 & 58.25 & 82.68 & 21101 & 1.29 \\  
		Qwen2(zero-shot) & 19.66 & 28.65 & 36.70 & 36.14 & 33590 & 2.68 \\   
		DeepSeek(zero-shot) &  &  &  &  & Out of Context Window &  \\  
		GLM4(compromise) & 80.33 & 85.39 & 92.09 & 90.45 & 39874 & 0.67 \\ 
		Qwen2(compromise) & 69.66 & 74.15 & 85.95 & 83.47 & 30268 & 1.06 \\  
		DeepSeek(compromise) & 71.34 & 77.52 & 86.37 & 85.15 & 60094 & 0.53 \\   
		GLM4(LR-SQL) & 94.38 & 97.19 & 96.91 & 97.85 & 45506 & 5.79 \\   
		Qwen2(LR-SQL) & 89.89 & 92.13 & 94.94 & 94.48 & 38374 & 5.33 \\ 
		DeepSeek(LR-SQL) & 75.28 & 88.76 & 87.07 & 91.43 & 75102 & 6.08 \\ 
		GLM4(DTS-SQL) & 96.63 & 97.75 & 97.75 & 98.17 & 78772 & 1.82 \\   
		Qwen2(DTS-SQL) & 92.70 & 94.38 & 97.08 & 96.82 & 67342 & 3.14 \\ 
		DeepSeek(DTS-SQL) & -- & -- & -- & -- & Out of Context Window & -- \\   
		\bottomrule  
	\end{tabular}   
\end{table*}  
Our experimental framework incorporates three recently released open-source models: ChatGLM4 9B\cite{glm2024chatglm}, Qwen2 7B\cite{yang2024qwen2}, and DeepSeek 7B\cite{bi2024deepseek}. The DeepSeek model is limited to a context length of 4096 tokens, restricting its capability to manage conversational inputs up to this length. Conversely, both ChatGLM4 9B and Qwen2 7B demonstrate the capacity to handle context lengths surpassing surpass a hundred thousand tokens. All experimental procedures were executed on a solitary A800-80G GPU, leveraging the LORA method for supervised fine-tuning. The LORA parameters were configured with R=64, Alpha=32, and a learning rate of 1e-5.
\subsection{Experimental model and hyperparameter settings}
Our method include two distinct models: a model for table prediction and a model for SQL query generation. 
The table prediction model is evaluated using four key metrics: Total Accuracy, Filtered Accuracy, Average Precision, and Average Recall \cite{pourreza2024dts}. Total Accuracy measures exact table matching, where predicted tables align perfectly with query targets. Filtered Accuracy assesses table coverage, ensuring predicted tables include all query targets.
The SQL generation model employs two main metrics: Execution Accuracy (EX) and Logical Form Accuracy (EM)\cite{yu2018spider}. Execution Accuracy (EX) requires predicted SQL results to match standard SQL queries exactly. Logical Form Accuracy (EM) evaluates the similarity between generated and annotated SQL.

\subsection{Results assessment}
\subsubsection{Dataset construction}

The Spider dataset originally contains training and validation sets that lack independent and identical distribution (i.i.d.). Nevertheless, models trained via supervised fine-tuning typically concentrate on specific private details and security, thereby aligning with i.i.d. conditions in real-world scenarios. As such, we opted to build our dataset upon the Spider training set.
By utilizing the Spider training set, we created an extensive database from original 164 database, filtering out duplicate table names to curate a selection of 254 distinct tables randomly, forming Dataset 1. This dataset include 4,030 question-answer pairs, partitioned into training and validation sets at a 9:1 ratio. Specifically, the training set comprises 3,647 instances, while the validation set includes 383 instances, each containing a question, concerning tables, and the accurate SQL query. During our experiments, we observed that encoding information from 256 tables resulted in an exceptionally high token count, and the memory demands for parameter-efficient fine-tuning of a medium-scale language model, with a batchsize of 1, surpassed the capacity of a single 80GB GPU.
To evaluate the distinctions between our LR-SQL and existing techniques, we subsequently randomly sampled 172 tables from Dataset 1, generating 1,914 question-answer pairs. These pairs were also divided into training and validation sets at a 9:1 ratio for Dataset 2. In Dataset 2, the training set consists of 1,736 instances, and the validation set comprises 178 instances.

\subsubsection{Schema\_Link Model Evaluation}

To showcase the efficacy of LR-SQL, we conducted a comparison with three distinct approaches for predicting database tables on two datasets, In the table, we use "-" to represent cases where experimental results could not be obtained.

Zero-shot: the method transfers question-answering templates to the model during inference, stimulating the model's existing knowledge while enhancing its ability to follow instructions.

Compromise: A highly streamlined supervised fine-tuning strategy that simply correlates questions, table names, and corresponding answers.

DTS-SQL\cite{pourreza2024dts}: A methodology that establishes connections between questions, comprehensive database contents, and concerning table or column names.

\begin{table*}[htbp]  
	\centering  
	\caption{Experiment on the Impact of Using Slices with Different Granularities on GPU Memory and Table Prediction Accuracy During Training}  
	\begin{tabular}{@{}c@{\hspace{4pt}}c@{\hspace{4pt}}c@{\hspace{4pt}}c@{\hspace{4pt}}c@{\hspace{4pt}}c@{\hspace{4pt}}c@{}}  
		\toprule  
		\multicolumn{7}{c}{\textbf{Dataset1}} \\  
		\midrule  
		Method & \centering Total \newline Accuracy & Filtered \newline Accuracy & Average \newline Precision & Average \newline Recall & GPU usage (MB) & Inference time(s) \\  
		\midrule  
		GLM4($w=6$, $slice\_token$=2100) & 87.99 & 93.47 & 94.20 & 94.71 & 69599 & 5.40 \\  
		GLM4($w=8$, $slice\_token$=1600) & 91.38 & 94.26 & 95.50 & 95.76 & 57152 & 6.49 \\  
		GLM4($w=10$, $slice\_token$=1300) & 90.60 & 93.47 & 95.10 & 95.44 & 50429 & 7.89 \\  
		GLM4($w=12$, $slice\_token$=1100) & 87.46 & 91.38 & 94.76 & 94.51 & 45825 & 9.56 \\ 
		\midrule 
		\multicolumn{7}{c}{\textbf{Dataset2}} \\  
		\midrule  
		& \centering Total \newline Accuracy & Filtered \newline Accuracy & Average \newline Precision & Average \newline Recall & GPU usage (MB) & Inference time(s) \\  
		\midrule  
		GLM4($w=6$, $slice\_token$=1600) & 93.82 & 95.50 & 95.36 & 95.88 & 55059 & 4.49 \\  
		GLM4($w=8$, $slice\_token$=1100) & 94.38 & 97.19 & 96.91 & 97.85 & 45506 & 5.79 \\  
		GLM4($w=12$, $slice\_token$=800) & 92.13 & 95.51 & 95.17 & 96.25 & 39155 & 7.01 \\  
		GLM4($w=14$, $slice\_token$=700) & 92.13 & 93.82 & 95.53 & 95.13 & 38263 & 9.34 \\  
		\bottomrule  
	\end{tabular}   
\end{table*}

During our experiments, we configured the total slice count by adjusting $slice\_token$, represented as total(w) = total(\_w) to be 8 for both the GLM4 and Qwen2 models. For the DeepSeek model, we assigned the slice number as total(w) = total(\_w) = 11 on Dataset1 and total(w) = total(\_w) = 8 on Dataset2. To facilitate a more accurate comparison of memory usage during supervised fine-tuning, we set the training batchsize to 1 for the DTS-SQL approach and to 2 for both the compromise and LR-SQL during the training phase. The ultimate table prediction outcomes for each model, employing the different methods, are detailed in Tables 1. All models are trained using our proposed method for 2 epochs on Dataset1 and for 3 epochs on Dataset2.


\begin{table}[htbp]  
	\centering  
	\caption{CoT prediction effectiveness evaluation result}  
	\begin{tabular}{@{}c@{\hspace{3pt}}c@{\hspace{3pt}}c@{\hspace{3pt}}c@{\hspace{3pt}}c@{}}  
		\toprule  
		\multicolumn{5}{c}{\textbf{Dataset1}} \\  
		\midrule  
		Method & \centering Total \newline Accuracy & Filtered \newline Accuracy & Average \newline Precision & Average \newline Recall \\  
		\midrule  
		GLM4(no\_CoT) & 88.51 & 94.52 & 95.63 & 96.74 \\  
		GLM4(CoT\_Injection) & 91.38 & 94.26 & 95.50 & 95.76 \\  
		GLM4(CoT) & 87.99 & 91.38 & 94.77 & 94.50 \\  
		DeepSeek(no\_CoT) & 74.15 & 88.51 & 90.07 & 93.72 \\  
		DeepSeek(CoT\_Injection) & 79.37 & 86.16 & 91.31 & 91.23 \\  
		DeepSeek(CoT) & 75.20 & 85.90 & 91.07 & 92.46 \\  
		Qwen2(no\_CoT) & 79.90 & 83.03 & 92.15 & 91.19 \\  
		Qwen2(CoT\_Injection) & 84.07 & 87.21 & 92.78 & 91.92 \\  
		Qwen2(CoT) & 66.58 & 67.62 & 90.77 & 81.24 \\ 
		\midrule 
		\multicolumn{5}{c}{\textbf{Dataset2}} \\  
		\midrule  
		Method & \centering Total \newline Accuracy & Filtered \newline Accuracy & Average \newline Precision & Average \newline Recall \\  
		\midrule 
		GLM4(no\_CoT) & 91.57 & 96.62 & 95.79 & 97.47 \\  
		GLM4(CoT\_Injection) & 94.38 & 97.19 & 96.91 & 97.85 \\  
		GLM4(CoT) & 88.76 & 91.01 & 97.19 & 95.08 \\  
		DeepSeek(no\_CoT) & 73.03 & 88.20 & 87.68 & 92.36 \\  
		DeepSeek(CoT\_Injection) & 75.28 & 88.76 & 87.07 & 91.43 \\  
		DeepSeek(CoT) & 73.60 & 85.39 & 88.90 & 90.96 \\  
		Qwen2(no\_CoT) & 88.76 & 92.13 & 94.38 & 94.57 \\  
		Qwen2(CoT\_Injection) & 89.89 & 92.13 & 94.94 & 94.48 \\  
		Qwen2(CoT) & 76.97 & 77.53 & 92.79 & 86.42 \\  
		\bottomrule  	 
	\end{tabular}  
\end{table}

The results presented in Tables 1 and 2 reveal that the zero-shot method exhibits the poorest performance. Relying solely on evoking the model's inherent pre-trained knowledge to address complex Text2SQL tasks still leaves a gap compared to supervised fine-tuning methods. The compromise approach, which merely establishes a relationship between questions and table names but overlooks column names, can accomplish table prediction tasks with lower computational resources; however, due to the lack of information, its performance remains inadequate in large-scale, complex Text2SQL tasks. Conversely, DTS-SQL showcases optimal performance across both datasets by seamlessly establishing direct associations among answers, databases, and concerning tables in one fell swoop. However, during the supervised fine-tuning process, the encoding of vast database tokens frequently necessitates substantial GPU memory usage, posing the risk of exceeding the current limitations of GPU capabilities, ultimately resulting in out-of-memory (OOM) errors. As evidenced by the findings in Dataset 1, this challenge underscores the importance of optimizing resource allocation to mitigate such issues. It is noteworthy that, compared to DTS-SQL, LR-SQL method achieves similar results while significantly reducing the GPU memory requirements.

\subsubsection{Evaluation of the Relationship between $slice\_token$, GPU Memory, and Performance}
In LR-SQL, the size of the GPU memory required for training is adjusted by regulating the $slice\_token$. Different size of  $slice\_token$ means encapsulating varying amounts of table information in the slice. Consequently, employing different size of $slice\_token$ also influences the inference performance of the model. Table 3 demonstrates the relationship between the constructed $slice\_token$ of varying granularities during the training of the GLM4 model, which exhibited the best performance, and the ultimate prediction accuracy.

The experimental outcomes suggest that our table prediction model is capable of dynamically modulating the memory demands during the supervised fine-tuning phase of the table name prediction task through the adjustment of $slice\_token$, all while maintaining comparable levels of the Filtered Accuracy metric.

\subsubsection{CoT prediction effectiveness evaluation}
The assessment of step-by-step prediction efficacy contrasts the ultimate results achieved by the task methodologies formulated through (6) depicted as 'no\_CoT' and (7) depicted as 'CoT\_Injection'. In order to validate that the enhancements credited to the approach delineated in (7) are not solely due to variations in inference templates, we examined an alternative strategy: during the training phase, the method omits the tables selected in prior steps, while during the inference phase, it integrates cues regarding the already selected. The aforementioned approach is depicted as 'CoT'. The number of slices for different models was configured consistently across the two datasets, in alignment with the settings used in section 2).

This experimental outcomes derived from the two datasets indicate that integrating data on the previously chosen tables during the supervised fine-tuning process can improve the model's table exact match rate to some degree, although this may lead to a minor reduction in the model's table coverage prediction proficiency. Upon contrasting the impacts of the 'CoT\_Injection' and 'COT' methodologies among diverse models, it becomes apparent that the enhanced performance observed with the 'CoT\_Injection' method is due to the way the tasks are formulated, and not merely a result of variations in inference templates.

\subsubsection{Experiment on Evaluating Slice Size during Inference}
During the inference process, the selection of varying $\_w$ sizes exerts an influence not only on the number of inference iterations but also on the accuracy of the model's predictions. As $slice\_token$ decreases, each slice incorporates a more substantial quantity of table information, consequently resulting in a greater number of encoded tokens. The relationship between the size of $\_w$, represented as $slice\_token\_inference$, at inference time and the size of $slice\_token$ during training is thoroughly examined and clarified through the subsequent experiments.The number of slices for different models was configured consistently across the two datasets, in alignment with the settings used in section 2) .

\begin{table*}[htbp]  
	\centering  
	\caption{Results of constructing slices with different granularities during inference on the final table prediction accuracy and inference time.}  
	\begin{tabular}{@{}c@{\hspace{4pt}}c@{\hspace{4pt}}c@{\hspace{4pt}}c@{\hspace{4pt}}c@{\hspace{4pt}}c@{}}  
		\toprule  
		\multicolumn{6}{c}{\textbf{Dataset1}} \\  
		\midrule  
		Method & \centering Total \newline Accuracy & Filtered \newline Accuracy & Average \newline Precision & Average \newline Recall & Inference time(s) \\  
		\midrule  
		GLM4(total($\_w$)=8) & 91.38 & 94.52 & 95.63 & 96.74 & 6.49 \\  
		GLM4(total($\_w$)=5) & 83.02 & 88.51 & 92.96 & 92.56 & 4.57 \\  
		GLM4(total($\_w$)=4) & 81.98 & 86.94 & 92.59 & 91.78 & 3.90 \\  
		DeepSeek(total($\_w$)=11) & 79.37 & 86.16 & 91.31 & 91.23 & 8.82 \\  
		DeepSeek(total($\_w$)=8) & 69.45 & 78.32 & 86.45 & 86.77 & 6.84 \\  
		DeepSeek(total($\_w$)=7) & 71.28 & 80.16 & 88.59 & 88.40 & 5.93 \\  
		Qwen(total($\_w$)=8)  & 84.07 & 87.21 & 92.78 & 91.92 & 5.49 \\  
		Qwen(total($\_w$)=5)  & 70.49 & 72.06 & 85.33 & 80.47 & 3.60 \\  
		Qwen(total($\_w$)=4)  & 68.15 & 69.19 & 82.13 & 76.91 & 3.08 \\ 
		\midrule  
		\multicolumn{6}{c}{\textbf{Dataset2}} \\  
		\midrule  
		Method & \centering Total \newline Accuracy & Filtered \newline Accuracy & Average \newline Precision & Average \newline Recall & Inference time(s) \\  
		\midrule  
		GLM4(total($\_w$)=8) & 94.38 & 97.19 & 96.91 & 97.85 & 5.79 \\  
		GLM4(total($\_w$)=5) & 88.7 & 91.57 & 95.13 & 94.42 & 3.91 \\  
		GLM4(total($\_w$)=4) & 83.70 & 86.51 & 93.63 & 92.13 & 3.30 \\  
		DeepSeek(total($\_w$)=8) & 79.37 & 86.16 & 91.31 & 91.23 & 6.08 \\  
		DeepSeek(total($\_w$)=5) & 69.66 & 76.97 & 87.27 & 85.81 & 4.34 \\  
		DeepSeek(total($\_w$)=4) & 74.15 & 78.65 & 89.33 & 85.77 & 3.87 \\  
		Qwen(total($\_w$)=8) & 89.89 & 92.13 & 94.94 & 94.48 & 4.99 \\  
		Qwen(total($\_w$)=5)  & 80.90 & 84.27 & 92.64 & 89.84 & 3.25 \\  
		Qwen(total($\_w$)=4)  & 79.21 & 82.02 & 91.95 & 87.78 & 2.72 \\  
		\bottomrule 
	\end{tabular}  
\end{table*}

The experimental results demonstrate that although setting a larger granularity for \_w can reduce the number of inference steps, the model exhibits better predictive performance when the value of $slice\_token\_inference$ during inference becomes increasingly similar to that during training.

\subsubsection{SQL Generation Model Evaluation}
The primary goal of Text2SQL is to transform user queries into structured query language statements. Following the application of the schema\_Link model, it becomes essential to deploy an SQL Generation model to produce the final SQL query. Our methodology aims to minimize the GPU memory required for supervised fine-tuning in the entire Text2SQL process. Therefore, we construct our SQL Generation model using Equation (4), which necessitates only minimal computational resources for supervised fine-tuning. We utilize Total\_GPU\_usage\_T to illustrate the maximum GPU memory required throughout the entire workflow, AVG\_inference\_time to represent the average inference time for a single data instance, and EX and EM to evaluate the quality of the generated SQL.
In the evaluation of SQL generation method, we benchmark our approach against the following methods:

Zero-shot: This technique harnesses prompt templates to activate the model's inferential abilities in SQL generation tasks.

DTS-SQL\cite{pourreza2024dts}: This strategy divides the Text2SQL task into two components: predicting the relevant tables and generating the SQL query.

DB\_GPT\cite{xue2023db}: This method directly maps the relationships between user questions, databases, and the resultant SQL statements.

\begin{table}[htbp]  
	\centering  
	\caption{The final results of using different supervised fine-tuning methods for Text2SQL}  
	\begin{tabular}{@{}lcccc@{}}  
		\toprule  
		\multicolumn{5}{c}{\textbf{Dataset1}} \\  
		\midrule  
		Method & EX & EM & Total\_GPU\_usage(MB) & Inference\_time(s) \\  
		\midrule  
		LR-SQL & 85.9 & 73.9 & 57152 & 4.53 + 6.49 \\  
		Zero-shot & 42.3 & 11.7 & 23048 & 2.87 \\  
		DTS-SQL & --  & --  & OOM & --  \\  
		DB\_GPT & --  & --  & OOM & --  \\
		\midrule  
		\multicolumn{5}{c}{\textbf{Dataset2}} \\  
		\midrule  
		Method & EX & EM & Total\_GPU\_usage(MB) & Inference\_time(s) \\  
		\midrule  
		LR-SQL & 84.8 & 77.0 & 45506 & 4.24 + 5.79 \\  
		Zero-shot & 41.0 & 14.0 & 20924 & 2.232 \\  
		DTS-SQL & 85.4 & 75.3 & 78772 & 4.32 + 1.82 \\  
		DB\_GPT & 83.1 & 80.9 & 80498 & 2.66 \\  
		\bottomrule  
	\end{tabular}  
\end{table}   

Based on the experimental results from LR-SQL and DTS-SQL, it can be observed that in large-scale and complex Text2SQL tasks, designing a table prediction model combined with training the SQL Generation model using (4) can reduce the information processing complexity of the SQL generation model, thereby enhancing the Execution Accuracy to a certain extent. Furthermore, LR-SQL is capable of decreasing GPU memory consumption throughout the entire Text2SQL pipeline while achieving performance comparable to existing approaches.

\section{Conclusion}
 We propose LR-SQL, which redesigns the objectives of supervised fine-tuning for the Schema\_link task by breaking down the database into slices for learning knowledge under limited GPU memory. The model can effectively learns the reasoning patterns of Chain-of-Thought and enhances the model's ability to perceive the entire database from partial slices during the inference phase. Ultimately, our method maintains performance close to the baseline while effectively reducing GPU memory consumption during fine-tuning. Furthermore, our approach can collaborate with QLORA to further decrease GPU memory usage during supervised fine-tuning.

\textbf{Future work}
 In our research, to effectively carry out the supervised fine-tuning of Text2SQL tasks under GPU memory limitations, we decompose the extensive database tables into several slices, facilitating batch-based prediction. While this strategy adeptly allows the model to retain a comprehensive understanding of the associations between queries and database tables, it does result in an increased number of inference steps during the model's operation. Nevertheless, our experimental findings demonstrate that the resulting inference latency stays within tolerable limits when juxtaposed with contemporary approaches. Hence, in our next step, we plan to investigate methods to reduce the overall time required for the LR-SQL inference process, while maintaining its robust performance.

\bibliographystyle{IEEEtran}  
\bibliography{references}

\end{document}